# Spin-to-charge conversion in orthorhombic RhSi topological semimetal crystalline thin films

Surya N. Panda[1*], Qun Yang[2], Darius Pohl[3], Hua Lv[1], Iñigo Robredo[1], Rebeca Ibarra[1], Alexander Tahn[3], Bernd Rellinghaus[3], Yan Sun[4], Binghai Yan[5], Anastasios Markou[1,6], Edouard Lesne[1*], and Claudia Felser[1*]

[1]Max Planck Institute for Chemical Physics of Solids, Nöthnitzer Str. 40, 01187 Dresden, Germany

[2]College of Letters and Science, University of California, Los Angeles, California 90095, USA

[3]Dresden Center for Nanoanalysis (DCN), Center for Advancing Electronics Dresden (CFAED), TUD Dresden University of Technology, D-01062 Dresden, Germany

[4]Institute of Metal Research, Chinese Academy of Science, Shenyang, China

[5]Department of Condensed Matter Physics, Weizmann Institute of Science, Rehovot, 7610001, Israel

[6]Physics Department, University of Ioannina, 45110 Ioannina, Greece

*surya.panda@cpfs.mpg.de; edouard.lesne@cpfs.mpg.de; claudia.felser@cpfs.mpg.de

## Abstract

The rise of non-magnetic topological semimetals, which provide a promising platform for observing and controlling various spin-orbit effects, has led to significant advancements in the field of topological spintronics. RhSi exists in two distinct polymorphs: cubic and orthorhombic crystal structures. The noncentrosymmetric B20 cubic structure has been extensively studied in the bulk for hosting unconventional multifold fermions. In contrast, the orthorhombic structure, which crystallizes in the Pnma space group (No. 62), remains less explored and belongs to the family of topological Dirac semimetals. In this work, we investigate the structural, magnetic, and electrical properties of RhSi textured-epitaxial films grown on Si(111) substrates, which crystallize in the orthorhombic structure. We investigate the efficiency of pure spin current transport across RhSi/permalloy interfaces and the subsequent spin-to-charge current conversion via inverse spin Hall effect measurements. The experimentally determined spin Hall conductivity in orthorhombic RhSi reaches a maximum value of $126\,\frac{\hbar}{e}(\Omega.\,cm)^{-1}$ at 10 K, which aligns reasonably well with first-principles calculations that attribute the spin Hall effect in RhSi to the spin Berry curvature mechanism. Additionally, we demonstrate the ability to achieve a sizable spin-mixing conductance (34.7 nm$^{-2}$) and an exceptionally high interfacial spin transparency of 88% in this heterostructure, underlining its potential for spin-orbit torque switching applications. Overall, this study broadens the scope of topological spintronics, emphasizing the controlled interfacial spin-transport processes and subsequent spin-to-charge conversion in a previously unexplored topological Dirac semimetal RhSi/ferromagnet heterostructure.

## Introduction

Topological semimetals are a novel class of quantum materials characterized by topologically non-trivial band structures[1], hosting Dirac[2], Weyl[3], or unconventional multifold fermions [4], which have been the focus of extensive research in recent years. These semimetals provide a promising alternative for efficient spin current generation in spin-orbit torque (SOT) devices due to their intrinsically large spin Hall and orbital Hall conductivities, which originate from Berry curvature effects[5-11]. Within the broader category of topological semimetals, transition metal-metalloid compounds (*e.g.*, RhSi, CoSi, PdGa, PtAl) have attracted significant attention for their ability to host unconventional chiral multifold fermions[12-14] when crystallized in the noncentrosymmetric cubic B20 structure with a chiral $P2_13$ space group (No. 198). This multifold fermionic behavior emerges from topologically protected band crossings (with four- and six-fold degeneracies) at certain high-symmetry points in the Brillouin zone, giving rise to several remarkable effects that are otherwise absent in other topological quantum materials[15-19].

RhSi is particularly notable because it also functions as a topological Dirac semimetal when it crystallizes in its orthorhombic rather than cubic structure. Geller *et al.* initially reported the cubic B20 crystal structure in RhSi-like transition metal-metalloid composite structures[20]. However, subsequent investigations by Schubert *et al.*[21] revealed that these compounds can also adopt a B31-type orthorhombic crystal structure with a Pnma space group (No. 62). Moreover, Mozaffari *et al.*[22] uncovered topological characteristics in the electronic band structure of orthorhombic RhSi, including a symmetry-protected Dirac nodal line and symmetry-enforced Dirac nodes at the S-point of the Brillouin zone. These Dirac nodes, which can be revealed by external magnetic fields, lead to anomalies in magnetic torque[22-24]. Despite these intriguing properties, driven by the topology of its electronic band structure, orthorhombic RhSi has not yet been explored for its potential in spin-charge current interconversion or in the broader field of topological spintronics. Among transition metal-metalloid Weyl semimetals (*e.g.*, RhSi, CoSi, PdGa, PtGa, PtAl), spin-to-charge interconversion has only been studied in polycrystalline cubic B20 CoSi thin films, using spin Hall magnetoresistance and harmonic Hall voltage measurement techniques[25]. However, detailed investigations to elucidate the magnitude of spin-mixing conductance, interfacial spin transparency, and the spin Hall angle in these B20-based heterostructures along with their temperature dependence are still lacking. Such studies would shed light on the microscopic mechanisms of spin relaxation and spin-to-charge interconversion.

In this work, we investigate non-magnetic topological semimetal RhSi/Ni$_{81}$Fe$_{19}$ (permalloy, hereafter Py) thin film heterostructures, grown via magnetron sputtering, using a combined ferromagnetic resonance (FMR)-driven spin-pumping generation and inverse spin Hall effect detection method. We experimentally determine the temperature dependence of the spin Hall angle ($\theta_{SH}$), spin diffusion length ($\lambda_{sd}$), and spin Hall conductivity ($\sigma_{SH}$) in RhSi. Transmission electron microscopy and X-ray diffraction reveal that RhSi crystallizes in an orthorhombic structure with an achiral centrosymmetric Pnma space group (No. 62), as depicted in Figure 1(a), which contradicts a previous report[26]. We observe that lowering the temperature significantly enhances $\theta_{SH}$, $\sigma_{SH}$, and $\lambda_{sd}$, while also affecting spin-mixing conductance and interfacial spin transparency. From the temperature dependence of the effective Gilbert damping, we conclude that intra-band conductivity-like scattering contributions dominate magnetization damping in Py thin films, while the enhanced Gilbert damping in RhSi/Py heterostructures is attributed to the spin-pumping mechanism. Furthermore, we examine the relative contributions of extrinsic mechanisms, such as spin-memory loss and two-magnon scattering, to the spin-pumping effect in RhSi/Py heterostructures.

## Results and discussions

### Samples characterization: X-ray diffraction (XRD)

The magnetron sputtering growth process of RhSi films and RhSi/Py heterostructures is detailed in the Methods section. The structure, crystallinity, and epitaxial relationship between the substrate and the sputter-deposited RhSi thin films were characterized using XRD measurements. Figure 1(b) shows $2\theta$-$\omega$ XRD scans of RhSi films with varying thicknesses ($t_{RhSi}$), each with a 3 nm-thick Si capping layer. In addition to the (111) and (222) reflections from the (111)-Si substrate, all samples exhibit the (200) and (400) reflections of orthorhombic B31-RhSi at $2\theta \approx$ 32.2° and 67.2°, respectively, indicating a preferred (100)-oriented growth of the RhSi films. We also observe a low-intensity broad peak at $2\theta \approx$ 46.4°, which corresponds to the (211) Bragg diffraction peak of orthorhombic RhSi. This finding contrasts with a previous report, which discussed RhSi films grown under similar conditions on Si(111) substrates as exhibiting a cubic B20-type structure[26]. The interpretation of the XRD data is supported by transmission electron microscopy (TEM) refinements of the structure of our films, as discussed later. Furthermore, the confirmation of textured-epitaxial growth in these RhSi thin films is evidenced by the asymmetric

XRD $\varphi$-scan (depicted in Figure 1(c)), which displays a six-fold periodicity of the {202} family of planes that coincides in $\varphi$ with the three-fold periodic {220} Bragg family of planes in the Si substrate. This suggests the presence of twin domains in the RhSi films, further confirmed by TEM analysis.

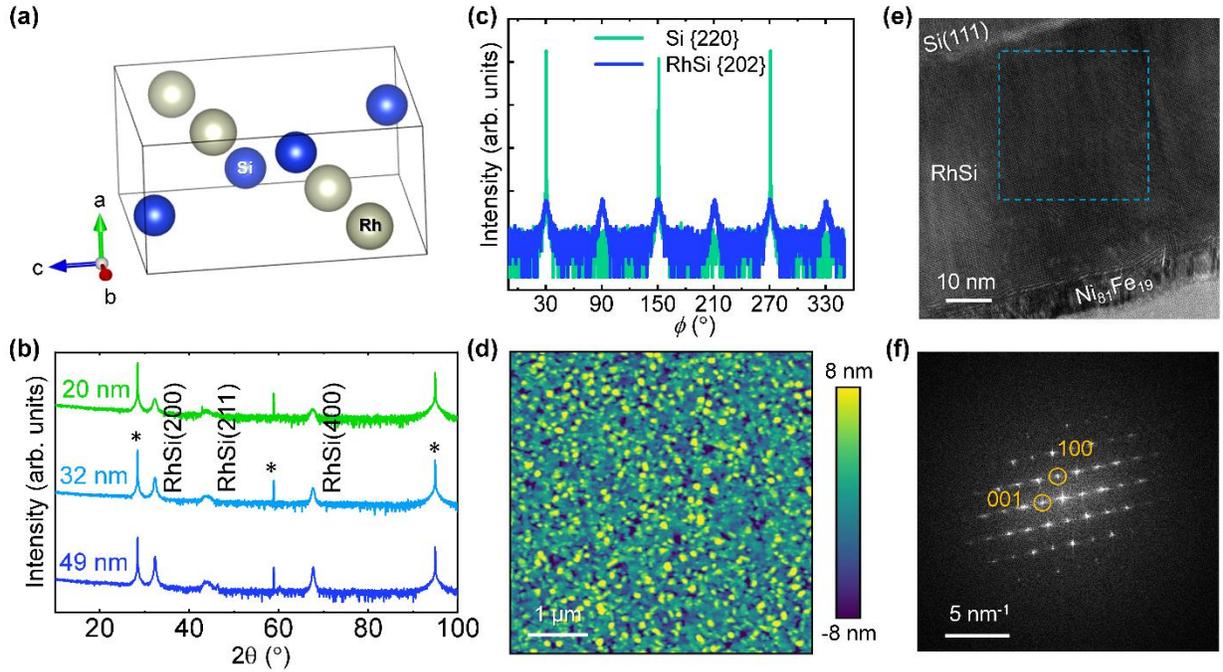

**Figure 1:** (a) Illustration of RhSi crystal structure that belongs to space group Pnma (No. 62). Grey and blue spheres denote Rh and Si atoms, respectively. (b) $2\theta$-$\omega$ XRD-pattern of RhSi thin films of different thicknesses grown on Si(111) single-crystal substrates (whose Bragg diffraction peaks are labeled by asterisks). (c) $\phi$-scan patterns of {220} family of planes from a 49 nm-thick RhSi film and {120} family of planes from the Si substrate (d) AFM-image of an uncapped 25 nm-thick RhSi film showing surface topography. Color bar encodes height. (e) High-resolution TEM image of a 49 nm-thick RhSi/Py(6.6 nm) heterostructure acquired along the [010] zone axis of the RhSi film. Blue dashed square marks the area for fast Fourier transform analysis in panel (f). (f) FFT of RhSi film with identified reflections consistent with lattice spacings of orthorhombic space group Pnma (No. 62).

**Samples characterization: Atomic force microscopy (AFM)**

We conducted surface topography mapping of the RhSi films using AFM on replicas of films with nominally identical thicknesses as those studied in this research, but without a Si capping layer. AFM imaging was conducted immediately after the films were removed from the deposition chamber to avoid potential surface degradation from prolonged exposure to ambient air. In Figure 1(d), an AFM-image of a 25 nm-thick RhSi film is presented, revealing an average root-mean-square roughness of 2.7 nm. Notably, the roughness was observed to increase slightly with increasing RhSi thickness. However, across the thickness range investigated in this study, no evidence of large-scale non-uniformity, discontinuity, or dislocations was detected.

**Samples characterization: Transmission electron microscopy (TEM)**

To further support the XRD analysis of the crystalline structure of RhSi, we performed high-resolution TEM (HRTEM) imaging of a 49 nm-thick RhSi film interfaced with a permalloy overlayer (see Methods for details). Figure 1(e) shows the resulting HRTEM image, focusing on a contoured region of interest within the RhSi film, for which a fast Fourier transform (FFT) analysis has been post-processed. The FFT image, presented in Figure 1(f), confirms that RhSi grows textured in its orthorhombic crystal structure (space group No. 62), with the [100] direction (*i.e.*, *a*-axis) aligned parallel to the [111] normal of the Si substrate. The detailed analysis of adjacent grains (see Supplementary Figure S1) reveals that they grow with either the [010] or [001] in-plane direction aligned parallel to the fixed [1$\bar{1}$0] zone axis of the Si substrate. This confirms the presence of twin domains in the RhSi films, with the *b*-axis and *c*-axis of the orthorhombic B31 structure lying within the plane of the film. Additionally, this allows us to determine the RhSi film's lattice parameters $a = 5.53$ Å, $b = 2.90$ Å and $c = 6.49$ Å.

**Samples characterization: DC electrical transport**

We also conducted temperature-dependent measurements of the longitudinal resistivity of RhSi films using a van der Pauw configuration. Across all thicknesses investigated, metallic behavior was observed, characterized by a continuous decrease in longitudinal resistivity as the temperature decreased from 320 K to 2 K. We found residual resistivity values as low as 55 μΩ.cm at 2 K, with minimal variation for films thicker than 20 nm (see Supplementary Figure S2). Notably, films of RhSi thinner than 19 nm exhibited higher resistivities and degraded metallic behavior, possibly due to surface and grain boundary scattering in the ultra-thin limit[27]. Therefore, we restricted our investigation to RhSi films thicker than 19 nm, which establishes an estimated thickness threshold for assessing the intrinsic electronic and spin-transport properties of these films.

**FMR-driven spin-pumping and inverse spin Hall measurements**

To harness the unique properties of non-magnetic topological semimetals in spin-based devices, it is necessary to investigate the efficiency of pure spin current generation and its subsequent transport across the interface with a ferromagnet (FM) [28]. Pure spin currents, which involve the flow of spins without any net charge flow, are critical for developing energy-efficient spin-based electronics, as they alleviate the limitations of charge-based devices, such as Joule heating and stray Oersted fields. Among the various mechanisms for generating spin currents [29], the "spin-pumping" phenomenon [30, 31] is particularly effective at generating pure spin currents at non-magnetic (NM)/FM interfaces, as it avoids the well-known impedance mismatch problem [32, 33]. In the FMR-driven spin-pumping process, the precession of magnetization in the FM layer induces a finite electrochemical potential at the NM/FM interface, caused by the asymmetric accumulation of majority and minority spins. This non-equilibrium spin accumulation generates a pure spin current that diffuses into the NM layer, where it can undergo spin-to-charge current conversion, as spin angular momentum is not conserved. This diffusive spin current provides an additional channel for damping the out-of-equilibrium magnetization driven by FMR, typically leading to an increase in the FM layer's Gilbert damping. The efficiency of the spin-pumping mechanism is primarily determined by the spin-mixing conductance (denoted $g_{\uparrow\downarrow}$) [34], while the magnitude of the injected spin current across the NM/FM interface is controlled by the interfacial spin transparency (denoted $\eta$) [35].

The FMR-driven spin-pumping approach has been extensively used to evaluate the primary spin-to-charge interconversion mechanisms and corresponding efficiencies [36], such as the inverse spin Hall effect (ISHE) and the associated spin Hall angle, and the inverse spin/orbital Rashba-Edelstein effect and corresponding Edelstein lengths. These processes convert a longitudinal spin current into a transverse electrical current [37-38]. In the NM layer, the charge current resulting from ISHE produces a measurable voltage corresponding to the electromotive force, $E_{ISHE}$, which follows the relation [39]: $E_{ISHE} \propto (\theta_{SH} \cdot g_{\uparrow\downarrow})$, where $\theta_{SH}$ is the spin Hall angle of the NM material, and $g_{\uparrow\downarrow}$ is the spin-mixing conductance at the NM/FM interface. For pure spin current-based device applications, it is necessary not only to identify NM materials with a high spin Hall angle but also to engineer NM/FM heterointerfaces with high interfacial spin transparency and spin-mixing conductance.

In this study, the magnetization dynamics and the resulting Gilbert damping were measured using a NanOsc Instruments cryo-FMR setup (see Methods section). A typical FMR spectrum for a RhSi(49 nm)/Py(6.6 nm) sample measured at various excitation frequencies is shown in Figure 2(a) for the range of 4–20 GHz. The collected FMR spectra represent the field-derivative of the imaginary part of the dynamic magnetic susceptibility $\chi$ as a function of the applied field ($\mu_0 H$). To determine the values of resonance field ($H_{res}$) and linewidth ($\Delta H$) from the FMR spectra, the data were fitted using the following formula[36]:

$$\frac{d\chi''}{dH}(H) = -K_{abs}\frac{(\Delta H)^2 - 4(H-H_{res})^2}{[4(H-H_{res})^2 + (\Delta H)^2]^2} + K_{dis}\frac{4\Delta H(H-H_{res})}{[4(H-H_{res})^2 + (\Delta H)^2]^2}, \quad (1)$$

where $K_{abs}$ and $K_{dis}$ are coefficients of the symmetric Lorentzian and antisymmetric components, corresponding to the absorption and dispersion contributions to the magnetic susceptibility, respectively. The effective saturation magnetization ($M_{eff}$) and anisotropy field ($H_k$) are obtained from the dispersion relation of the resonance frequency vs. field $f_{res}(H_{res})$, as described by Kittel's formula[40]:

$$f_{res} = \frac{\gamma \mu_0}{2\pi}\sqrt{(H_{res} + H_k)(H_{res} + H_k + 4\pi M_{eff})}, \quad (2)$$

where $\gamma = \frac{g\mu_B}{\hbar}$ is the gyromagnetic ratio, $\mu_B$ and $\hbar$ are the Bohr magneton and reduced Planck constant, respectively. $g$ is the Landé $g$-factor for the ferromagnetic layer. Using this fitting, $M_{eff}$, $g$, and $H_k$ are determined as fitting parameters. Figure 2(b) shows the dispersion relation $f_{res}$ vs. $H_{res}$ for the RhSi(49 nm)/Py(6.6 nm) and Py(6.6 nm) samples. The open symbols show the experimental data, while the solid lines represent the fit using Equation (2). From these fits, we obtain values of $M_{eff} \approx 630 \pm 8$ kA m$^{-1}$ in the presence of RhSi, and $(692 \pm 10)$ kA m$^{-1}$ in its absence. This decrease in $M_{eff}$ suggests a change in interfacial anisotropy, possibly due to modified interfacial spin-orbit interaction strength in the presence of RhSi[41]. The effective electron $g$-factor is found to be 2.11± 0.01 for both samples, a typical value for permalloy[42]. In the absence and presence of RhSi the anisotropy field ($H_k$) value is 2.1 and 1.1 mT, respectively.

The overall magnetization dynamics in the presence of spin pumping can be described by a modified version of the Landau–Lifshitz–Gilbert equation, given by[30, 31]:

$$\frac{d\bm{m}}{dt} = -\gamma(m \times H_{\text{eff}}) + \alpha_{\text{int}}\left(m \times \frac{dm}{dt}\right) + \frac{\gamma}{VM_s}\left(I_s^{\text{pump}} - I_s^{\text{back}}\right), \qquad (3)$$

where $V$ is the volume of the FM, $M_s$ is its saturation magnetization, $I_s^{\text{pump}}$ is the spin current injected into the NM layer, and $I_s^{\text{back}}$ is the backflow spin current returning to the FM layer. $m$ denotes the magnetization ($\vec{M}$) unit vector, i.e., $m = \frac{\vec{M}}{|\vec{M}|}$. The net spin angular momentum flow across the NM/FM interface is determined by the balance between $I_s^{\text{pump}}$ and $I_s^{\text{back}}$.

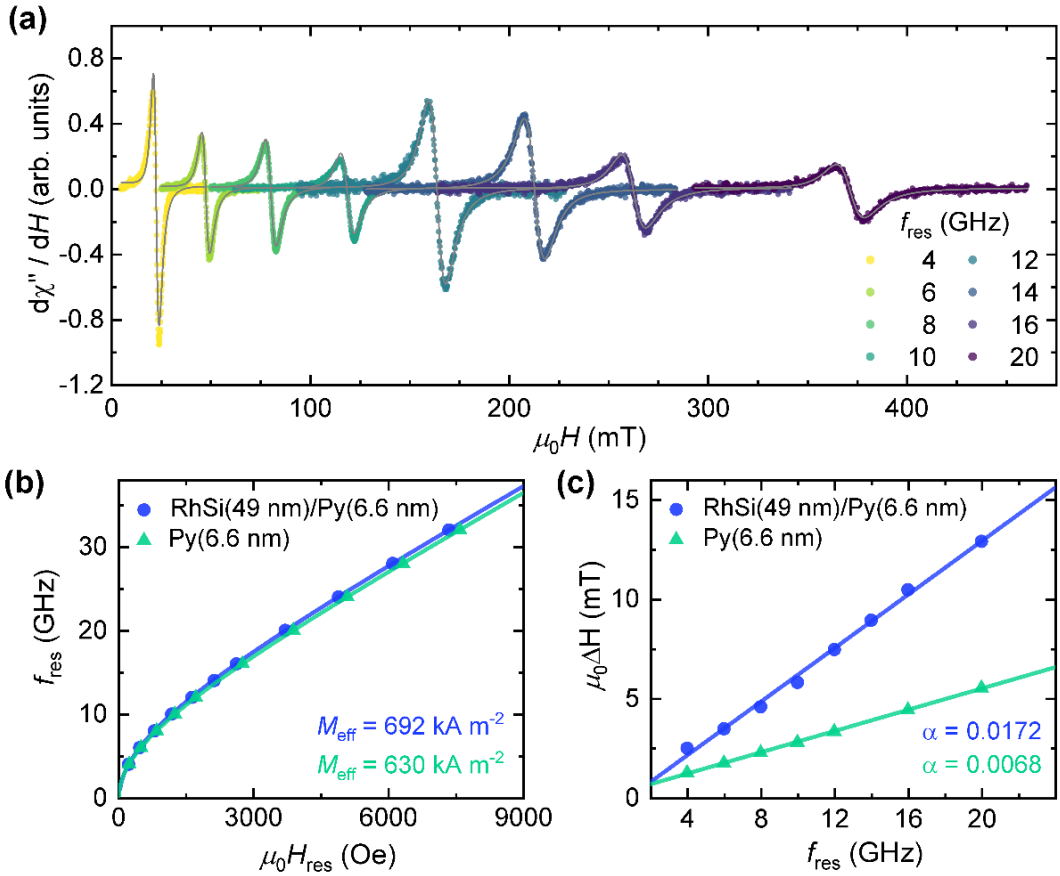

**Figure 2:** (a) FMR spectra of RhSi(49 nm)/Py(6.6 nm) sample measured at different resonant excitation frequencies ($f_{\text{res}}$). The symbols correspond to experimental data and solid lines are the best fit using Eq. (1). (b) Field dependence of resonance frequency for RhSi(49 nm)/Py(6.6 nm) and Py(6.6 nm) samples. Solid lines are fitting results using Kittel's formula, Eq. (2), which is parameterized by $M_{\text{eff}}$. (c) Linewidth vs. resonance frequency of FMR response. Solid lines indicate best fit using Eq. (4), whose slope corresponds to the Gilbert damping $\alpha$.

The Gilbert damping parameter ($\alpha$) is calculated by fitting the $f$-dependent $\Delta H$ using the relationship[43]:

$$\Delta H = \frac{4\pi}{\gamma} \alpha f_{\text{res}} + H_0 \qquad (4)$$

where $H_0$ denotes the frequency-independent linewidth broadening, which depends on magnetic inhomogeneities within the sample. The effective Gilbert damping parameter $\alpha$ obtained from this analysis includes contributions from intrinsic spin-pumping and extrinsic effects including spin-memory loss, two-magnon scattering, interfacial band hybridization, and magnetic proximity effects. Figure 2(c) shows the $\Delta H$ vs. $f_{\text{res}}$ dispersion relation for the RhSi(49.1 nm)/Py(6.6 nm) and Py(6.6 nm) samples. The linear behavior of $\Delta H(f_{\text{res}})$ implies good homogeneity in our samples. $H_0$ value is found to be less than 1 mT in both the presence and absence of the RhSi underlayer. The determined value of $\alpha$ for RhSi(49.1 nm)/Py(6.6 nm) is 0.0172 ± 0.0005, which is significantly higher than the Gilbert damping parameter for the Py(6.6 nm) reference sample 0.0068 ± 0.0004). This difference is consistent with the presence of an efficient spin-pumping effect, as discussed further. However, it is important to note that other extrinsic factors may also contribute to the observed enhancement in $\alpha$.

The transfer of spin angular momentum across the NM/FM interface, without any spin backflow effect, is characterized by the intrinsic spin-mixing conductance, denoted as $g_{\uparrow\downarrow}$. This intrinsic $g_{\uparrow\downarrow}$ represents the conductance properties of spin channels at the NM/FM interface when the NM layer thickness ($t_{\text{NM}}$) is much greater than the spin diffusion length ($\lambda_{\text{sd}}$), which represents the characteristic distance traveled by spin currents before dissipation in the NM layer. In the presence of spin backflow (where $t_{\text{NM}} \leq \lambda_{\text{sd}}$), spin transport through the interfacial spin channels is described by an effective spin-mixing conductance ($g_{\uparrow\downarrow}^{\text{eff}}$), which depends on the material properties, the interface, and the thickness of the NM layer. The flow of spin angular momentum across the interface exerts a damping-like torque on the magnetization, leading to an increase in the Gilbert damping parameter. The modulation of the Gilbert damping parameter, $\Delta\alpha = (\alpha_{\text{NM/FM}} - \alpha_{\text{FM}})$, can be related to the spin-mixing conductance as follows[36, 44]:

$$\Delta\alpha = \frac{g\mu_B}{4\pi t_{\text{FM}} M_s} g_{\uparrow\downarrow}^{\text{eff}}, \qquad (5)$$

where $t_{FM}$ is the thickness of the FM layer. In the presence of the spin-pumping effect, $\alpha$ increases non-monotonically and saturates as the thickness of the NM layer increases. It is important to note that this description of the spin-pumping effect assumes a constant and thickness-independent resistivity of the NM layer. However, the effective Gilbert damping in NM/FM can be affected sensitively by changes in resistivity of either or both the NM and FM layers[45, 46]. In this study, we have taken care of maintaining the thickness and resistivity of the Py layer constant, and have carefully measured the temperature-dependent resistivity of our RhSi films across the entire thickness series. Therefore, for films thicker than 19 nm, as studied here, the observed enhancements in $\alpha$ primarily reflect the efficiency of the spin-pumping effect. Some studies[47] have indicated that thermal gradients in thin film heterostructures can lead to spin Seebeck effect and anomalous Nernst effect. However, in Py-based systems, and for spatially extended blanket films of several mm$^2$ studied here (see Methods), such prospective thermal gradients are very small[48], and the temperature elevation of the sample at resonance negligible (of the order of hundreds of mK). Thus, the aforementioned thermally-driven effects are not expected to contribute relevantly to the overall spin-pumping signal.

**ISHE measurements and determination of spin Hall conductivity**

In NM materials exhibiting a sizable SHE, an efficient spin-pumping effect is accompanied by a characteristic voltage drop, $V_{ISHE}$ due to the ISHE[37]. This effect causes the generated spin current density, $J_s$, to be converted into a transverse charge current density, $J_c$ (or a DC voltage in an open circuit), within the high SOC material. The efficiency of this interconversion mechanism is parametrized by the spin Hall angle $\theta_{SH} = \left(I_s/I_c\right)$. The schematic experimental geometry of the spin-pumping and combined ISHE measurement configuration is shown in Figure 3(a); see the Methods section for further details. The overall voltage drop (denoted $\Delta V$) is composed of both field-symmetric and antisymmetric components (see Supplementary Figure S3), which can be separated by fitting the experimental data with a combination of (anti)Lorentzian functions[49-50]:

$$\Delta V = V_{sym}\left[\frac{(\Delta H)^2}{(H-H_{res})^2+(\Delta H)^2}\right] + V_{asym}\left[\frac{\Delta H(H-H_{res})}{(H-H_{res})^2+(\Delta H)^2}\right]. \tag{6}$$

where $V_{sym}$ is attributed to the ISHE contribution[30, 31, 36], and $V_{asym}$ represents contributions from other extrinsic effects, such as rectification[51]. The generated charge current, $I_c$, within the RhSi layer is given by the ratio between $V_{ISHE}$ and the sheet resistance of the RhSi film, $I_c = V_{ISHE}/(\rho_{RhSi} \cdot t_{RhSi})$. Figure 3(b) displays the measured voltage drop minus an offset voltage ($\Delta V$), at 300 K for a series of RhSi films with various thicknesses (ranging from 20 to 49 nm), measured during FMR-spin pumping and ISHE experiments conducted at 8 GHz. Additionally, the temperature dependence of $\Delta V$ is shown in Figure 3(c), where the overall amplitude of $\Delta V$ monotonically increases as the system temperature decreases from 300 K to 10 K.

In Figure 3(d), we have plotted the RhSi thickness dependence of the ISHE-induced charge current $I_c$ at different temperatures. We observe that $I_c$ increases monotonically with increasing RhSi thickness and eventually saturates at higher thicknesses, in agreement with Eq. (7) under the ISHE mechanism. Additionally, the magnitude of $I_c$ systematically increases as temperature decreases, indicating enhanced spin-to-charge interconversion efficiency at lower temperatures in our RhSi thin films, as discussed in the next section.

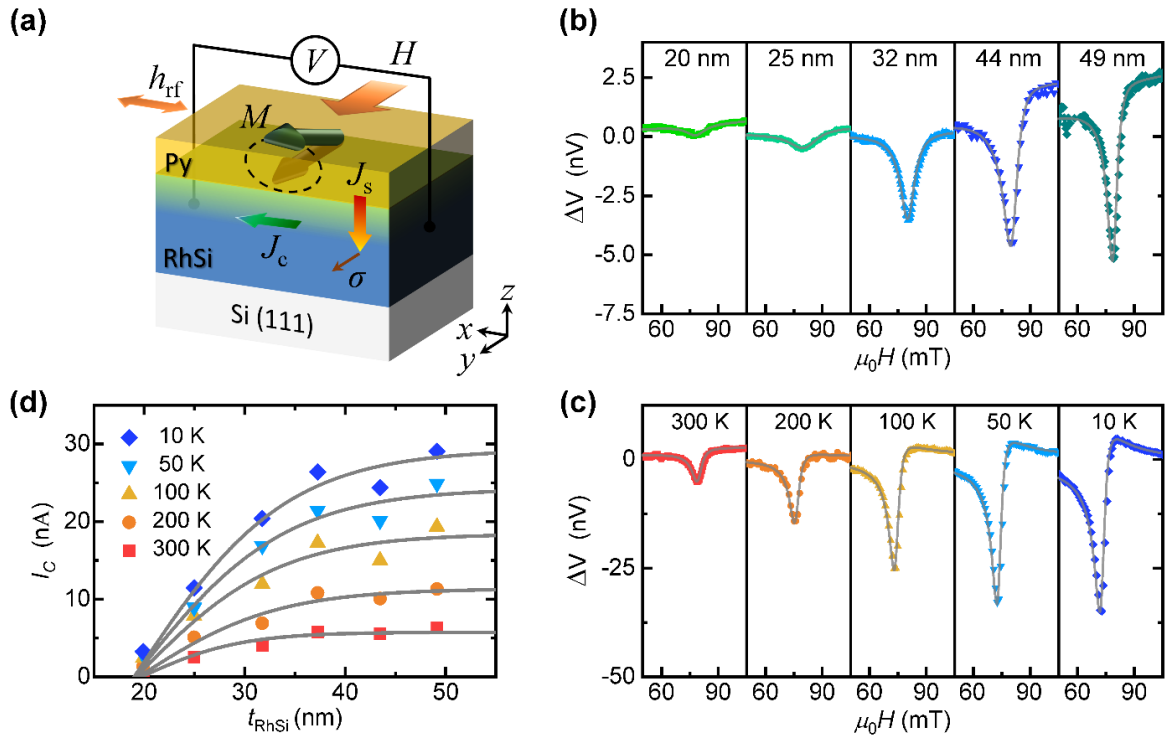

**Figure 3:** (a) Schematic of sample structure and measurement geometry for FMR-driven spin pumping and inverse spin Hall effect (ISHE) measurements. Voltage drop detected in FMR-spin pumping experiments at 8 GHz (b) for RhSi($t_{RhSi}$)/Py(6.6 nm) heterostructures at 300 K, and (c) at various temperatures (10–300 K) for a 49 nm-thick RhSi/Py(6.6 nm) sample. (d) RhSi-thickness-dependent modulation of ISHE-induced charge current fitted using Eq. (7) (solid red lines), at various temperatures.

The reported RhSi thickness-dependent $I_c$ can be used to extract the characteristic spin Hall angle ($\theta_{SH}$) and spin diffusion length ($\lambda_{sd}$) of RhSi, which are related by the following expression[52, 53]:

$$I_c = I_s \, w \, \theta_{SH}^{eff} \, \lambda_{sd} \cdot \tanh\left[\frac{t_{RhSi}}{2\lambda_{sd}}\right], \quad (7)$$

where $I_s$, the spin current generated by spin pumping, is given by:

$$I_s = \frac{2e}{\hbar}\left(\frac{g_{\uparrow\downarrow}^{eff}\hbar}{8\pi}\right)\left(\frac{\mu_0 h_{rf}\gamma}{\alpha}\right)^2 \left[\frac{\mu_0 M_{eff}\gamma + \sqrt{(\mu_0 M_{eff}\gamma)^2 + (4\pi f)^2}}{(\mu_0 M_{eff}\gamma)^2 + (4\pi f)^2}\right]. \quad (8)$$

$w$ is the width of the sample, and $h_{rf}$ is the magnitude of the radiofrequency (RF) magnetic field. According to the calibration (see Methods), $h_{rf} \approx 60$ µT. We denote $\theta_{SH}^{eff}$ the effective SH angle extracted using this formalism. As will be discussed later, the estimation of an intrinsic $\theta_{SH}$ value requires considering additional mechanisms (*e.g.*, interfacial spin transparency), which can affect its magnitude and cause $\theta_{SH}$ to differ from the measured $\theta_{SH}^{eff}$. The behavior of the measured $I_c$ vs. $t_{RhSi}$ is well described by Eq. (7) at all measurement temperatures, as shown by the solid red lines in Figure 3(d). The value of $g_{\uparrow\downarrow}^{eff}$, extracted from Eq. (5), is used in Eq. (8) to extract the magnitude of the spin current density $J_s = I_s / (t_{RhSi} \cdot w)$ at each temperature.

Figure 4(a) shows the variation of $J_s$ and $g_{\uparrow\downarrow}^{eff}$ with temperature. We observe that $J_s$ decreases whereas $g_{\uparrow\downarrow}^{eff}$ increases with increasing temperatures. The decrease in $J_s$ can be attributed to increased dissipation of accumulated spin densities at the interface due to both bulk and interfacial spin scattering, which also leads to the observed increase in $\alpha$[54]. The rise in $g_{\uparrow\downarrow}^{eff}$ with temperature signifies a higher probability of spin transfer and increased losses of spin angular momentum from the FM at elevated temperatures. This trend in $g_{\uparrow\downarrow}^{eff}$ also indicates that interfacial spin accumulation decreases with increasing temperature facilitating a more efficient spin transfer between Py and RhSi at higher temperatures. This trend in $g_{\uparrow\downarrow}^{eff}$ could be further attributed to the presence of some level of interfacial disorder (both structural and electronic) whose impact is more pronounced at lower temperature and which consequently reduces the spin-mixing conductance in our Py/RhSi samples.

Figure 4(b) shows the temperature dependence of $\theta_{SH}$ and $\lambda_{sd}$ extracted from the fitting of $I_c$ vs. $t_{RhSi}$ in Figure 3(d). We chose to display $\theta_{SH}$ as the measured effective spin Hall angle, corrected by the estimated value of the interfacial spin transparency $\eta$ (as discussed in the following section). Regardless, we observe that both $\theta_{SH}$ and $\lambda_{sd}$ decrease with increasing temperatures. The decrease in $\lambda_{sd}$ with temperature is more straightforward to understand and is expected when the Elliott-Yafet (EY) scattering mechanism dominates the spin relaxation process. Within the EY mechanism, the spin scattering probability scales with the momentum relaxation rate[55], leading to $\lambda_{sd} \propto (\rho_{RhSi})^{-1}$, where $\rho_{RhSi}$ is the longitudinal resistivity of the RhSi films. We measured $\rho_{RhSi}=$ 109.2 µΩ.cm (56.7 µΩ.cm) at 300 K (10 K), resulting in $\lambda_{sd} = 4.9$ nm (7.5 nm). This shorter $\lambda_{sd}$ at higher temperatures is also due to higher spin resistance ($\propto \frac{\rho_{RhSi}}{V}$), where $V$ is the volume in which the spin current diffuses, as predicted by the spin diffusion model[56]. The temperature dependence of $\theta_{SH}$ and the spin Hall conductivity (SHC), $\sigma_{SH}$, in RhSi clarifies the mechanisms contributing to the ISHE in our heterostructures. Notably, $\theta_{SH}$ decreases from 1.8% at 10 K to a moderate 0.6% at RT, a behavior similar to that observed in Au films[56], which are known to exhibit very low intrinsic SHC.

Furthermore, it is possible to estimate the SHC of RhSi films. Under the condition where the longitudinal resistivity $\rho_{RhSi} \gg \rho_{SH}$, the SHC is given by: $\sigma_{SH} = \frac{\theta_{SH}}{\rho_{RhSi}}\left(\frac{\hbar}{e}\right)$. The temperature dependence of $\sigma_{SH}$ is displayed in Figure 4(c). The value of $\sigma_{SH}$ decreases from $106 \frac{\hbar}{e}(\Omega.\text{cm})^{-1}$ at 10 K to $25 \frac{\hbar}{e}(\Omega.\text{cm})^{-1}$ as the temperature increases to 300 K, as shown in Figure 5(c). The total $\sigma_{SH}$, is characteristically the sum of intrinsic and extrinsic contributions, acting as parallel channels ($\sigma_{SH} = \sigma_{SH}^{int} + \sigma_{SH}^{ext}$). The intrinsic $\sigma_{SH}^{int}$, originating from the spin Berry curvature mechanism, is temperature-independent because it does not rely on the momentum relaxation time and thus is not expected to depend on $\rho_{RhSi}$. However, the extrinsic term $\sigma_{SH}^{ext}$ originates from various extrinsic scattering mechanisms, such as skew scattering at impurities and phonons, or side-jump scattering in highly resistive materials. This extrinsic contribution is expected to display a temperature dependence that reflects its multiple originating mechanisms. Although we cannot entirely rule out these contributions to the measured signal, assessing their individual contributions remains challenging. For these reasons, in a following section, we rather decide to focus the discussion on the intrinsic Berry curvature source of SHE in RhSi in the context of first-principles calculations.

**Role of interfacial spin transparency in spin transport across RhSi/Py interface**

The generation of a pure spin current through spin pumping does not necessarily ensure that all the spins accumulating at the NM/FM interface will successfully diffuse through the NM layer and undergo spin-to-charge conversion therein. Factors such as electronic band alignment, disorder, and intermixing at the NM/FM interface play a crucial role in determining the probability of spin transmission across the interface. The concept of interfacial spin transparency, denoted $\eta$, accounts for all these effects, which influence whether electrons are reflected from the interface or transmitted during spin transport. In the diffusive spin-transport model, $\eta$ can be expressed as a function of $g_{\uparrow\downarrow}^{\text{eff}}$ and $\lambda_{sd}$ using the following relation[35,57]:

$$\eta = \frac{g_{\uparrow\downarrow}^{\text{eff}} \tanh\left(\frac{t_{\text{NM}}}{2\lambda_{\text{sd}}}\right)}{g_{\uparrow\downarrow}^{\text{eff}}\coth\left(\frac{t_{\text{NM}}}{\lambda_{\text{sd}}}\right)+\frac{h}{2\lambda_{\text{sd}}e^2\rho_{\text{NM}}}} \,, \tag{10}$$

where $\rho_{\text{NM}}$ is the resistivity of the NM layer, $h$ is Plank's constant, and $e$ is the elementary charge.

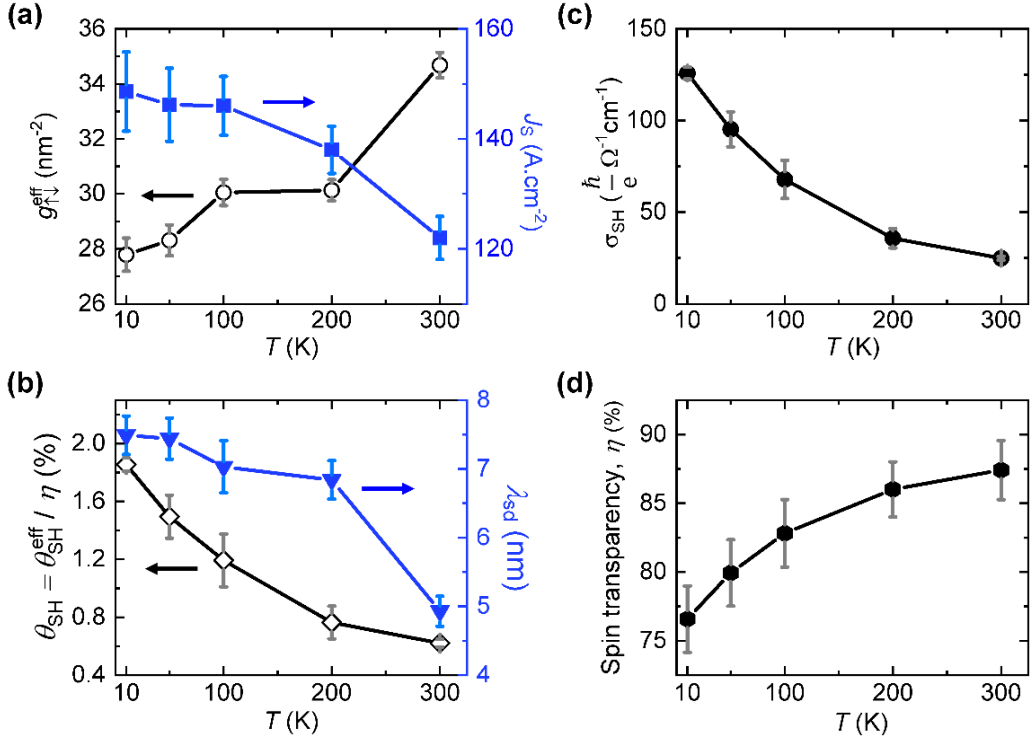

**Figure 4:** Temperature-dependent variation of (a) effective spin-mixing conductance (left axis) and generated charge current via spin pumping (right axis), (b) spin Hall angle (left) and spin diffusion length (right) of RhSi, (c) spin Hall conductivity of orthorhombic RhSi films, and (d) interfacial spin transparency of RhSi/Py heterostructures.

The magnitude of $\eta$ provides a qualitative measure of the spin transparency at the RhSi/Py interfaces. Recent studies suggest that in NM/FM heterostructures, both $\theta_{SH}$ and the strength of the SOT are dependent on the precise value of $\eta$ [35, 58, 59]. In Figure 4(d), we display the $\eta$-value for RhSi/Py heterointerfaces, which shows a monotonic increase with rising temperature. The spin transparency remains consistently high, varying from 77% at 10 K to 88% at 300 K. This increase in interfacial spin transparency $\eta$ may be attributed to improved electronic band matching at higher temperatures and the activation of thermally activated spin-transport mechanisms across the interface. Consequently, the spin transport becomes less sensitive to local disorder or imperfections that could otherwise hinder interfacial spin transmission.

We stress out that the specific temperature dependence of each spin-dependent quantity reported in Figure 4 cannot be easily attributed uniquely to one given microscopic mechanism, nor can we exclude some level of extrinsic contribution, as mentioned earlier. This includes spin-dependent scattering at defects and impurities and their rate of change due to temperature-dependent electron-phonon interactions or screening effects. Additional contributions due to thermal expansion (and their mismatch at interfaces), changes in lattice parameters and their impact on strain and the electronic band structure (of both RhSi and the spin-active interfacial region), as well as on the position of the Fermi level are beyond the scope of the present experimental investigation, but could warrant further attention.

**Electronic band structure and SHC calculations**

To gain insights into the intrinsic origin of spin-to-charge interconversion in RhSi, which is driven by the spin Berry curvature, we performed density functional theory (DFT) calculations using the Vienna Ab *initio* simulation package for orthorhombic RhSi (space group No. 62)[60, 61], with the experimentally determined lattice parameter values. Within the framework of the SHE, the externally applied electric field along the $\beta$-direction ($E_\beta$) and the induced spin current along the α-direction, with spin polarization along the $\gamma$-direction ($J_\alpha^\gamma$), are related by the SHC tensor ($\sigma_{\alpha\beta}^\gamma$) as $J_\alpha^\gamma = \sigma_{\alpha\beta}^\gamma E_\beta$. We reframe ($\alpha,\beta,\gamma$) to align with the natural coordinate system ($x$, $y$, $z$), constrained by the experimental geometry and the preferential *a*-axis texture of the RhSi films (see Figure 3(a)). In this system, the $z$-axis aligns with the *a*-axis of RhSi (which also corresponds to the direction of the spin current density, $J^z$), whereas the $x$- and $y$-axes lie within the film plane, corresponding to the *c*- and *b*-axis directions of RhSi, respectively. Thus, the SHC matrix elements $\sigma_{zx}^y$ and $\sigma_{zy}^x$ are the relevant components accessible through the experimental technique and geometry (see Table 1). We calculated $\sigma_{zx}^y$ and $\sigma_{zy}^x$ to be -65 and 190 $\frac{\hbar}{e}(\Omega.\text{cm})^{-1}$, respectively, at the Fermi level.

Notably, the experimental value of $\sigma_{SH}$ at 10 K (126 $\frac{\hbar}{e}(\Omega.\text{cm})^{-1}$) closely matches the average of these two inequivalent SHC tensor components. This agreement is consistent with the presence of twin domains oriented in-plane along either the [001] or [010] principal axis directions of RhSi, as supported by XRD and TEM characterizations (see Figure 1 and Supplementary Figure S1). Beyond the quantitative agreement between experiment and theory, the sign and magnitude of the measured SHC are fully consistent with expectations from an intrinsic spin Berry curvature mechanism for a twinned (100)-textured RhSi orthorhombic crystal.

**Table 1:** Calculated spin Hall conductivity of orthorhombic RhSi at the Fermi level. SHC magnitude in units of $\frac{\hbar}{e}(\Omega.\text{cm})^{-1}$. z-direction (x and y) corresponds to [100] axis (in-plane [001] and [010]) of RhSi.

| | $\sigma^x$ | $\sigma^y$ | $\sigma^z$ |
|---|---|---|---|
| SG 62<br>6 independent components | $\begin{pmatrix} 0 & 0 & 0 \\ 0 & 0 & \sigma_{yz}^x \\ 0 & \sigma_{zy}^x & 0 \end{pmatrix}$ | $\begin{pmatrix} 0 & 0 & \sigma_{xz}^y \\ 0 & 0 & 0 \\ \sigma_{zx}^y & 0 & 0 \end{pmatrix}$ | $\begin{pmatrix} 0 & \sigma_{xy}^z & 0 \\ \sigma_{yx}^z & 0 & 0 \\ 0 & 0 & 0 \end{pmatrix}$ |
| z [100] | $\begin{pmatrix} 0 & 0 & 0 \\ 0 & 0 & -190 \\ 0 & 190 & 0 \end{pmatrix}$ | $\begin{pmatrix} 0 & 0 & 61 \\ 0 & 0 & 0 \\ -65 & 0 & 0 \end{pmatrix}$ | $\begin{pmatrix} 0 & -166 & 0 \\ 185 & 0 & 0 \\ 0 & 0 & 0 \end{pmatrix}$ |

We calculated the *k*-resolved spin Berry curvature of the band structure and the energy-dependent SHC. As shown in Figures 5(a) and 5(b), the red and blue color-coded scales denote significant positive and negative contributions to the spin Berry curvature, respectively. Orthorhombic B31-RhSi exhibits topological features with multiple Dirac nodes and a symmetry-protected Dirac nodal line in its band structure[22]. Specifically, the band structure reveals multiple band crossings along high-symmetry (HS) lines, starting from the HS point *S* in its 3D Brillouin Zone (BZ). Additionally, symmetry-protected band crossings occur along the SX lines near the Fermi energy, leading to a symmetry-protected nodal line around the *S*-point on the $k_x = \pi$ plane. Moreover, the SY, TR, and RU HS lines are identified as fourfold-degenerate HS lines enforced by the crystal symmetries. As shown clearly in Figures 5(a) and 5(b), these fourfold-degenerate HS lines serve as sources of substantial spin Berry curvature distributions $\Omega^x_{zy}$ and $\Omega^y_{zx}$. Moreover, because the SHC depends on the Fermi level, it varies rapidly when the Fermi energy shifts by only a few tens of meV. Figures 5(c) and 5(d) illustrate that both $\sigma^x_{zy}$ and $\sigma^y_{zx}$ exhibit peaks of 587 and -137 $\frac{\hbar}{e}(\Omega.\text{cm})^{-1}$, respectively, when the Fermi energy is positioned at approximately -0.1 eV, which is near the band crossing points along the SX lines. This energy-dependent analysis indicates a potential strategy for enhancing the SHE in RhSi by fine-tuning the Fermi level position, which could be achieved through techniques such as epitaxial strain, external pressure, substitutional doping or electrostatic gating.

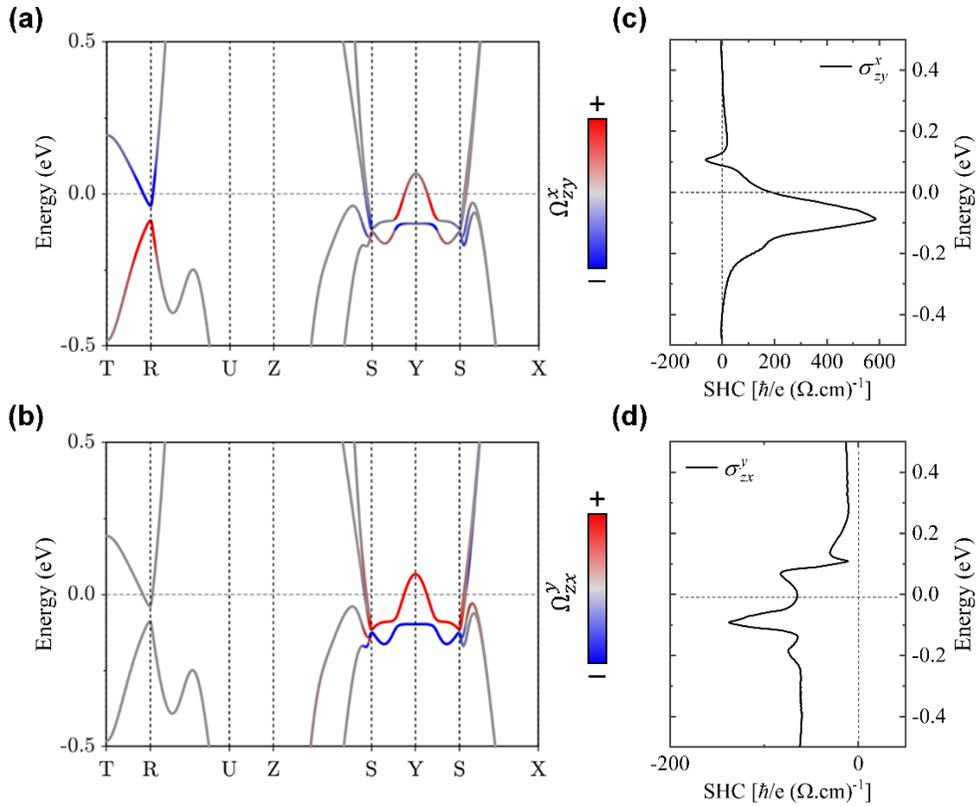

**Figure 5**: Spin Berry curvature resolved band structure along high-symmetry lines of the BZ for (a) $\Omega^x_{zy}$ and (b) $\Omega^y_{zx}$ of orthorhombic RhSi. Red (blue) denotes positive (negative) contributions. (c, d) Corresponding energy-dependent *k*-integrated SHC tensor elements accessible experimentally.

## Conclusions

In summary, we investigated spin pumping and the ISHE in textured-epitaxial (100)-RhSi topological semimetal thin films, grown on Si(111) substrates, which crystallized in the orthorhombic structure with the Pnma space group No. 62. We demonstrated that, in the presence of RhSi, the magnetization dynamics are dominated by the spin-pumping effect, which can be adjusted by changing the thickness of RhSi and Py. We evaluated the Gilbert damping parameter, spin-mixing conductance, interfacial spin transparency, spin Hall angle, and SHC for the RhSi/Py system by modeling the experimental results. Additionally, we achieved broad tunability of these parameters with system temperature and RhSi thickness. At the lowest accessible temperature (10 K), RhSi/Py heterostructures displayed a maximum spin Hall angle of 1.8% for RhSi, a spin-mixing conductance of 34.7 nm$^{-2}$, and an interfacial spin transparency of 88%. The substantial intrinsic SHC, along with the high interfacial spin transparency in RhSi/Py, suggest that orthorhombic RhSi is a promising material for pure spin current-based spintronics and spin-orbitronics applications. The values of $g_{\uparrow\downarrow}^{\text{eff}}$ and $\eta$ for the RhSi/Py interface are significantly higher than those of the widely used Pt/Py interface and compare positively with a number of other canonical (*e.g.*, Pt/YIG) and more exotic NM/FM heterostructures, such as: incorporating topological insulators, van der Waals layered transition metal dichalcogenides or Heusler alloys, and frequently studied in spintronic devices (as summarized in Table 2). This further confirms that RhSi serves as an excellent spin sink, resulting in a robust spin pumping effect. The temperature dependence of $\eta$, $\theta_{\text{SH}}$, $\sigma_{\text{SH}}$, $M_{\text{eff}}$, $\alpha$, $\lambda_{\text{sd}}$, and $g_{\uparrow\downarrow}$ provides essential guidance for selecting the appropriate material combination for spintronics devices with desired operating conditions. For spin-based device applications, these parameters can be further optimized by selecting different NM topological semimetals, engineering interfacial layers (*e.g.*, to tailor Rashba spin-orbit coupling[62], or enhance interfacial spin transparency), and adjusting their thicknesses, in conjunction with a chosen FM layer. Therefore, the parameter space for enhancing the functionalities and performance of topological semimetal/FM interfaces for spintronics remains largely unexplored.

Our findings may pave the way for the development of spin-orbitronics devices based on newly synthesized topological Dirac semimetals in thin film form, especially in applications such as SOT magnetic random-access memories, where $\eta$, $\theta_{\text{SH}}$, and $\alpha$ play a critical role in enhancing switching speed and overall efficiency. We believe that various strategies, such choosing specific combinations of FM and NM materials, electrostatic field effect, strain, or tuning materials' composition and Fermi level via chemical doping, could enhance the intrinsic SHC, improve interfacial spin current transport, as well as increase the overall efficiency of spin-to-charge current conversion. Furthermore, disentangling the contributions of the putative orbital Hall effect[63, 64] where finite spin-orbit interaction is not required from the overall transverse signal attributed to the SHE opens up new possibilities in the emerging field of orbitronics, potentially allowing the generation and propagation of orbital rather than spin information over distances far exceeding characteristic spin diffusion lengths[65].

**Table 2:** Comparison of the (effective) spin-mixing conductance ($g_{\uparrow\downarrow}$), spin diffusion length ($\lambda_{sd}$), spin Hall angle ($\theta_{SH}$) and interfacial spin transparency ($\eta$) of the RhSi/Py samples studied here with selected values at spin Hall material/FM interfaces taken from the literature. N.A. stands for data not available.

| Heterostructures | $g_{\uparrow\downarrow}$ (nm$^{-2}$) | $\lambda_{sd}$ (nm) | $\theta_{SH}$ (%) | $\eta$ |
|---|---|---|---|---|
| RhSi/Py (This work) | 34.7 | 7.5 | 1.8 | 0.88 |
| Pt/Py[37] | 15.2 | 1.4 | 19 | 0.25 |
| Pd/Py[66] | 14.0 | 5.8 | N.A. | N.A. |
| Au/YIG[67] | 2.7 | 60 | 8.4 | N.A. |
| Ag/YIG[67] | 5.2 | 700 | 0.68 | N.A. |
| W/YIG[67] | 4.5 | 2.1 | 14 | N.A. |
| Ta/YIG[67] | 5.4 | 1.9 | 7.1 | N.A. |
| Pt/YIG[67] | 6.9 | 7.3 | 10 | N.A. |
| Co$_2$MnGa/Py[68] | N.A. | 3.1 | -19 | N.A. |
| Bi$_2$Se$_3$/CoFeB[69] | 12 | 5.0 | 43 | N.A. |
| CoSi/CoFeB[25] | N.A. | 4.4 | 3.4 | N.A. |
| Mn$_3$Ga/CoFeB[70] | 38 | 1.0 | 31 | N.A. |
| Mn$_3$Au/CoFeB[71] | 8.8 | 1.6 | 22 | 0.61 |
| MoS$_2$/CoFeB[72] | 14.3 | 7.83 | N.A. | N.A. |
| Pt/Co$_2$FeAl$_{0.5}$Si$_{0.5}$[51] | 3.7 | 3.06 | 1.6 | N.A. |
| NbN/YIG[73] | 10 | 14 | -1.1 | N.A. |

# Methods

## A. Magnetron sputtering growth

Thin films of RhSi($t_{RhSi}$)/Py($t_{Py}$)/SiO$_x$(3 nm) were grown employing a BESTEC UHV magnetron sputtering system on (111)-oriented Si single-crystal substrates. The thickness of the RhSi layer was varied from 19 nm to 55 nm, and the Py layer thickness ranged from 6.6 nm to 30 nm. Before the deposition process began, the chamber was evacuated to a base pressure of less than $5 \times 10^{-9}$ mbar, with the process gas (Ar 5 N) pressure set to $3 \times 10^{-3}$ mbar. The target-to-substrate distance was maintained at 18.6 cm to ensure optimal sputtering conditions. To achieve spatial uniformity, the substrate was rotated at 24 rpm during deposition. The growth of (100)-oriented RhSi films was conducted at 680°C, with Rh and Si sources in confocal geometry operated at 15 W DC power and 80 W RF power, respectively. Following the RhSi deposition, the ferromagnetic Py layer was deposited at room temperature using a DC power of 40 W applied to a Ni$_{81}$Fe$_{19}$ target. Each sample was capped *in situ* with a protective Si layer to prevent oxidation of the metallic silicide or permalloy films. The deposition conditions were carefully optimized and maintained consistently for all samples.

## B. Characterizations

Symmetric and asymmetric X-ray diffraction (XRD) scans were carried out using a Panalytical XPert[3] XRD diffractometer with Cu K$\alpha_1$ radiation ($\lambda$ = 1.5406 Å). The electron density, interface roughness, and layer thicknesses were determined through X-ray reflectivity measurements. Surface topography was analyzed using atomic force microscopy (AFM) with an MFP-3D Origin microscope from Asylum Research (Oxford Instruments). TEM was conducted with a JEOL JEM F200, operated at an acceleration voltage of 200 kV and equipped with a GATAN OneView CMOS camera for fast imaging. Local EDS analysis was performed using a dual 100 mm$^2$ window-less silicon drift detector.

For longitudinal resistivity measurements, we employed a Physical Properties Measurement System (PPMS) Quantum Design cryostat and ultrasonically bonded aluminum wires to the corners of the square samples in a van der Pauw geometry. Frequency-dependent FMR measurements were conducted with a NanOsc FMR setup, integrated with a Quantum Design PPMS. The sample, typically 3 mm x 5 mm in size, was positioned in a flip-chip configuration atop a 200 μm wide coplanar waveguide (CPW). FMR measurements were conducted across frequencies ranging from 4 to 20 GHz (and up to 40 GHz), spanning temperatures from 10 K to 300 K, in an in-plane geometry. As per the careful RF power *vs.* frequency calibration provided by the FMR spectrometer instrument provider (NanOsc), the RF field magnitude is estimated to be 60 μT and nearly frequency-independent in the 4 to 20 GHz range used for this study. The magnetic field was swept across a predefined range while the frequency remained constant. To measure the ISHE, electrical contacts were established on opposite sides of the long edge of the sample using silver paste and platinum wires, and the ISH voltage was measured using a Keithley 2182A nanovoltmeter. We have also carefully verified and optimized the response of our cryogenic NanOsc FMR setup using standard $Ni_{81}Fe_{19}$/Pt thin film heterostructures, which yield values of spin-mixing conductance, spin Hall angle and spin diffusion length within the characteristically reported values in the literature for this specific benchmark system for spin-pumping/ISHE experiments.

## C. First-principles calculations

The exchange-correlation potential was described using the generalized gradient approximation, following the Perdew-Burke-Ernzerhof parametrization scheme[74]. A *k*-point grid of 8 × 8 × 8 was employed, and the total energy convergence criterion was set to $10^{-6}$ eV. From DFT calculations, we projected the ab initio DFT Bloch wavefunctions into highly symmetric atomic-orbital-like Wannier functions using a full-potential local-orbital minimum-basis code (FPLO)[75]. This allowed us to generate the corresponding tight-binding (TB) model Hamiltonian, which fully respects the symmetry of the materials under study. In the context of SHE, the applied electric field along the *β*-direction ($E_\beta$) and the induced spin current along the α-direction with spin polarization along the *γ*-direction ($J_\alpha^\gamma$) are related by the SHC tensor ($\sigma_{\alpha\beta}^\gamma$) as $J_\alpha^\gamma = \sigma_{\alpha\beta}^\gamma E_\beta$. Using the obtained TB Hamiltonian, the intrinsic SHC tensor $\sigma_{\alpha\beta}^\gamma$ was calculated via the Kubo formula:

$$\sigma_{\alpha\beta}^\gamma = e\hbar \int_{BZ} \frac{d\mathbf{k}}{(2\pi)^3} \sum_n f_{n\mathbf{k}} \Omega_{n,\alpha\beta}^{\hat{S}_\gamma}(\mathbf{k}), \qquad (11)$$

$$\Omega_{n,\alpha\beta}^{\hat{S}_\gamma}(\mathbf{k}) = -2\text{Im} \sum_{m \neq n} \frac{\langle n(\mathbf{k})|\hat{j}_\alpha^{\hat{S}_\gamma}|m(\mathbf{k})\rangle \langle m(\mathbf{k})|\hat{v}_\beta|n(\mathbf{k})\rangle}{(E_n(\mathbf{k}) - E_m(\mathbf{k}))^2}. \qquad (12)$$

where $\hat{j}_\alpha^{\hat{S}_\gamma} = \frac{1}{2}\{\hat{v}_\alpha, \hat{S}_\gamma\}$ is the conventional spin current operator, and $\Omega_{n,\alpha\beta}^{\hat{S}_\gamma}(\mathbf{k})$ represents the spin Berry curvature. $\hat{S}_\gamma$ is the spin operator, $E_n(\mathbf{k})$ is the eigenvalue for the $n_{th}$ eigenstate $|n(\mathbf{k})\rangle$ at momentum **k**, and $\hat{v}_{\alpha(\beta)}$ is the α(β) component of the band velocity operator defined by $\hat{v}_{\alpha(\beta)} = \frac{1}{\hbar}\frac{\partial \hat{H}(\mathbf{k})}{\partial k_{\alpha(\beta)}}$, and $f_{n\mathbf{k}}$ is the Fermi-Dirac distribution function. For the integration in Eq. (11), a uniform 240×240×240 k-grid was used to perform the k-space summation.

We aligned the z-axis with the crystallographic a-axis, corresponding to the experimental setup where the spin current flows along the z-axis (crystallographic *a*-axis) and the charge current flows along either the x- or y-axis. Orthorhombic RhSi belongs to the space group Pnma, which is defined by the twofold screw rotations $\{2_{001} | \frac{1}{2}, 0, \frac{1}{2}\}$, $\{2_{010} | 0, \frac{1}{2}, 0\}$ and inversion symmetry $\{-1 | 0\}$. These symmetries combine to give three screw axis operations $\{2_{001} | \frac{1}{2}, 0, \frac{1}{2}\}$, $\{2_{100} | \frac{1}{2}, \frac{1}{2}, \frac{1}{2}\}$, and $\{2_{010} | 0, \frac{1}{2}, 0\}$, as well as three glide plane operations $\{m_{001} | \frac{1}{2}, 0, \frac{1}{2}\}$, $\{m_{100} | \frac{1}{2}, \frac{1}{2}, \frac{1}{2}\}$, and $\{m_{010} | 0, \frac{1}{2}, 0\}$. The material also exhibits time-reversal symmetry. Constrained by these symmetry operations, many SHC tensor elements are forced to zero, leaving only six independent non-zero elements, as shown in Table 1.

**Supporting Information**

RhSi TEM grain analysis; longitudinal resistivity of RhSi crystalline thin films; temperature- and thickness-dependence of Gilbert damping parameter and saturation magnetization; Permalloy thickness-dependent FMR; symmetric and antisymmetric components of ISH voltage measurements.

**Corresponding Authors**

Surya N. Panda: surya.panda@cpfs.mpg.de

Edouard Lesne: edouard.lesne@cpfs.mpg.de

Claudia Felser: claudia.felser@cpfs.mpg.de

**Notes**

The authors declare no competing financial interest.

## Acknowledgments

This work was supported by the Horizon 2020 FETPROAC Project No. SKYTOP-824123, "Skyrmion-Topological Insulator and Weyl Semimetal Technology" and the Sächsische Aufbaubank - Förderbank (SAB) project "Topologische Spintronic: CMOS-kompatible Materialien aus der B20-Familie (TOP20)," Cluster No. 4188. Q.Y. gratefully acknowledges funding from the Otto Hahn Award of the Max Planck Society, under funding No. 85931. We acknowledge the use of the facilities in the Dresden Center for Nanoanalysis (DCN) at the Technische Universität Dresden.

# – Supporting Information –

# Spin-to-charge conversion in orthorhombic RhSi crystalline thin films

Surya N. Panda[1*], Qun Yang[2], Darius Pohl[3], Hua Lv[1], Iñigo Robredo[1], Rebeca Ibarra[1], Alexander Tahn[3], Bernd Rellinghaus[3], Yan Sun[4], Binghai Yan[5], Anastasios Markou[1,6], Edouard Lesne[1*], and Claudia Felser[1*]

[1]Max Planck Institute for Chemical Physics of Solids, Nöthnitzer Str. 40, 01187 Dresden, Germany
[2]College of Letters and Science, University of California, Los Angeles, California 90095, USA
[3]Dresden Center for Nanoanalysis (DCN), Center for Advancing Electronics Dresden (CFAED), TUD Dresden University of Technology, D-01062 Dresden, Germany
[4]Institute of Metal Research, Chinese Academy of Science, Shenyang, Liaoning, 110016, China
[5]Department of Condensed Matter Physics, Weizmann Institute of Science, Rehovot, 7610001, Israel
[6]Physics Department, University of Ioannina, 45110 Ioannina, Greece

* surya.panda@cpfs.mpg.de; edouard.lesne@cpfs.mpg.de; claudia.felser@cpfs.mpg.de

## I. Grain analysis using transmission electron microscopy

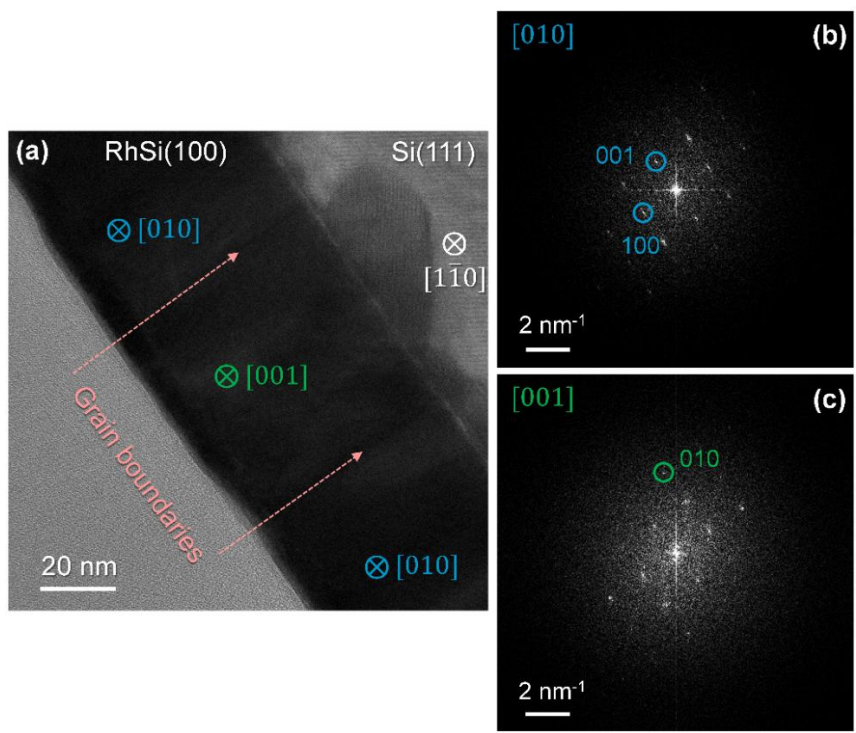

**Figure S1:** (a) Cross-section HRTEM image of RhSi (same as in Fig. 1e) showing differently oriented crystalline grains and their boundaries (marked by arrows). Local fast Fourier transform showing adjacent grains possessing (b) [010] and (c) [001] in-plane crystallographic orientations

The textured growth of the RhSi film was investigated by HRTEM imaging. Figure S1(a) shows a cross section image of three grains in the RhSi film. All grains grow with the [100] direction parallel to the [111] substrate normal, verified by local FFTs (see Figures S1(b) and S1(c)). The adjacent grains have [010] and [001] zones axis with the (001) and (010) planes parallel to the substrate surface. The grains have a size of roughly 60 nm with straight grain boundaries from the RhSi/Si interface to the RhSi/Py interface.

.

**II. Longitudinal resistivity of RhSi crystalline thin films**

The temperature-dependent measurements of the longitudinal resistivity ($\rho_{xx}$) were conducted on thin films of RhSi of various thicknesses. These measurements were performed using a van der Pauw geometry within a PPMS (Physical Property Measurement System) cryostat provided by Quantum Design. RhSi samples capped *in situ* with Si (forming a naturally oxidized ~ 3nm-thick $SiO_x$ capping layer), but without a permalloy overlayer were deposited in order to assess their intrinsic electrical transport properties. The four corners of square-shaped samples were contacted via ultrasonic wedge-bonding technique, using aluminium wires. In Figure S2, we present the longitudinal resistivity versus temperature, $\rho_{xx}(T)$, for RhSi(*t*)/SiO$_x$(3 nm) films of various thicknesses. All the films exhibit a metallic behavior, characterized by a continuous decrease in $\rho_{xx}(T)$ as the temperature is lowered from 320 K down to 2 K. The residual resistivity values ($\rho_{xx,0}$), at 2 K, are comprised between 54.6 µΩ.cm for the thicker film and 69 µΩ.cm for the thinner one. The corresponding range of residual resistivity ratios (RRR) is 1.7-2.0. Furthermore, we note that the resistivity did not exhibit a saturation behaviour at low *T*, implying that within the temperature range under investigation, $\rho_{xx,0}$ was not primarily influenced by extrinsic scattering mechanisms. Specifically, this suggests that factors such as grain boundaries, impurities, interfaces, or crystallographic defects whose presence would typically lead to higher residual resistivity are not the dominant contributors to the low *T* resistivity in our RhSi films.

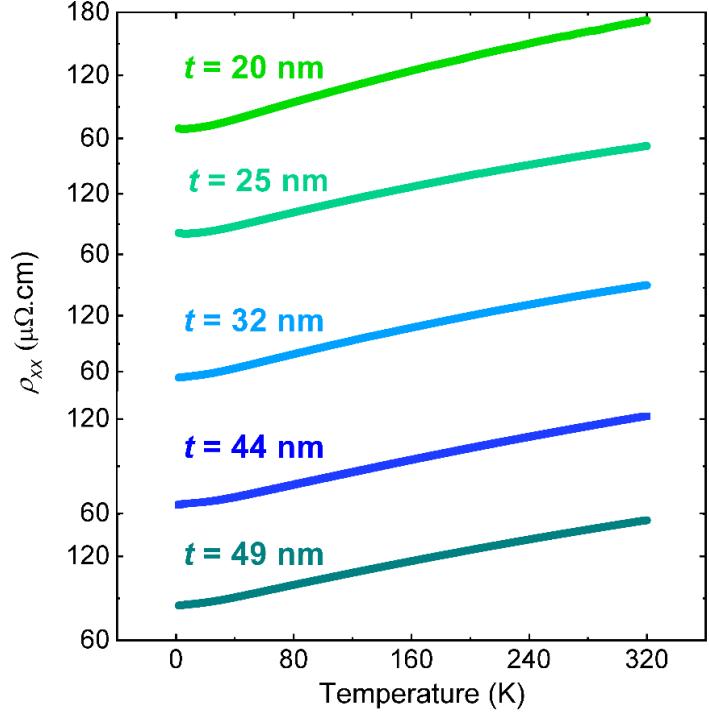

**Figure S2:** Longitudinal resistivity as a function of temperature, $\rho_{xx}(T)$, for RhSi($t$) crystalline thin films (capped with 3 nm-thick SiO$_x$).

### III. Symmetric and antisymmetric components in ISHE measurements

The measured ISHE voltage is plotted against the external magnetic field in Figure S3(a) for the RhSi(49.1 nm)/Py(6.6 nm) sample at 8 GHz frequency. Here, $V_{\text{sym}}$ and $V_{\text{asym}}$ components are shown by solid red and magenta lines, extracted from the fit using Eq 6. A significantly higher value of $V_{\text{sym}}$ in comparison to $V_{\text{asym}}$ indicates that spin pumping is the dominant mechanism in the enhancement of $\alpha$ in RhSi/Py heterostructures. The reversal of the sign of $V_{\text{ISHE}}$ is observed by reversing the magnetic field (in Figure S3(b)), which excludes that this ISHE signal is produced by a possible thermoelectric effect induced by the ferromagnetic resonance absorption.

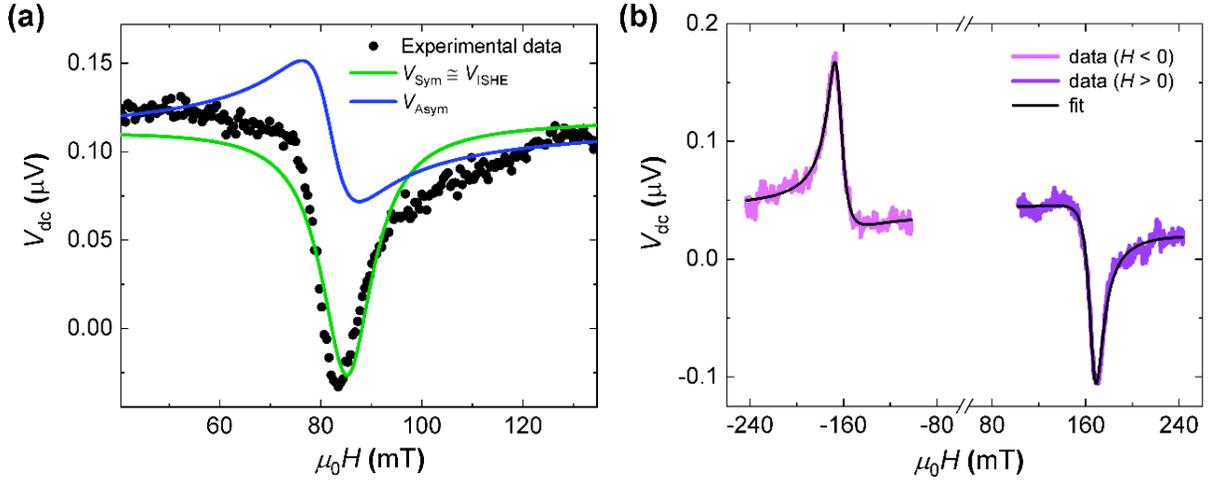

**Figure S3:** (a) Measured dc-voltage ($V_{dc}$) *vs.* applied magnetic field for the RhSi(49 nm)/Py(6.6 nm) sample at 8 GHz. The solid green corresponds to the field-symmetric contribution which is identified as the ISHE signal. The solid blue line corresponds to the field-antisymmetric contribution. (b) Measured $V_{dc}$ upon in-plane magnetic field reversal. The solid black lines are the fit using Eq. 6 of the main manuscript.

## IV. Temperature- and thickness-dependence of Gilbert damping parameter and saturation magnetization

In Figure S4, we present the modulation in effective saturation magnetization ($M_{eff}$) and Gilbert damping parameter ($\alpha$) with the thickness of RhSi and system temperature. Our findings reveal that $\alpha$ maintains a relatively constant across varying RhSi thicknesses, indicating the absence of spin angular momentum backflow through spin pumping. Moreover, this underscores that the spin diffusion length ($\lambda_{sd}$) of RhSi is significantly less than 19 nm. Additionally, we observe a monotonic decrease in $\alpha$ with decreasing temperature in Py thin films, as shown in Figure S5(a). However, in RhSi/Py heterostructures, $\alpha$ increases with temperature (as displayed in Figure S5(b)). This temperature dependence of $\alpha$ in magnetic thin films can be modeled as [S1, S2]:

$$\alpha(T) = \alpha_{\text{int}}(T) + \alpha_{\text{SP}}, \tag{S1}$$

$$\text{with } \alpha_{\text{int}}(T) = \alpha_{\text{intra}} \frac{\sigma(T)}{\sigma(300\ K)} + \alpha_{\text{inter}} \frac{\rho(T)}{\rho(300\ K)}. \tag{S2}$$

Here, $\alpha_{\text{int}}$ denote the intrinsic Gilbert-like damping of the FM and $\alpha_{\text{SP}}$ the additional damping contribution arising from spin pumping into the adjacent NM layer. $\alpha_{\text{intra}}$ and $\alpha_{\text{inter}}$ relate to the intra-band and inter-band contributions, respectively, which are found to be dependent on the conductivity ($\sigma$) and resistivity ($\rho$) of the sample. $\alpha_{SP}$ is characteristically inversely proportional to the temperature while the $\alpha_{int}$ can have both direct or inverse dependence on temperature depending upon the leading mechanism of spin-dependent scattering. In the RhSi/Py heterostructures, the increase in $\alpha$ with temperature indicates a dominant intra-band conductivity-like scattering contribution, which is proportional to the momentum relaxation time. Higher temperatures enhance the localization of interactions between magnons and conduction electrons in these heterostructures. In contrast, when the RhSi underlayer is absent, the decrease in $\alpha$ with temperature suggests that inter-band resistivity-like contributions play a predominant role in spin-dependent scattering processes during magnetization dynamics.

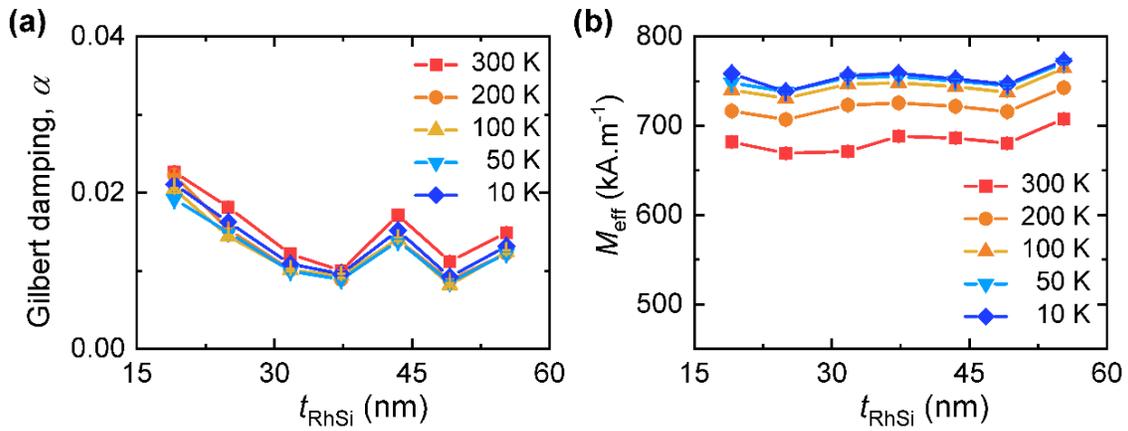

**Figure S4:** Temperature- and RhSi-thickness-dependent modulation of (a) Gilbert damping parameter $\alpha$, and (b) effective saturation magnetization $M_{eff}$.

Furthermore, our observations indicate that $M_{eff}$ remains nearly constant across different RhSi thicknesses (see Figure S4(b)). This implies that interfacial anisotropy remains consistent throughout the range of RhSi thicknesses examined, exerting only a minor influence on the modification of magnetization dynamics within these heterostructures. Both in the presence and absence of RhSi underlayer (shown in Figure S5), $M_{eff}$ decreases monotonically with temperature. This decrease in $M_{eff}$ with temperature can be directly correlated to the Bloch's law ($M_{eff}(T) = M_{eff}(0)(1 - AT^{3/2})$), where $M_{eff}(0)$ and $B$ represent the saturation magnetization at T = 0 K and a parameter related to the exchange stiffness constant, respectively [S3].

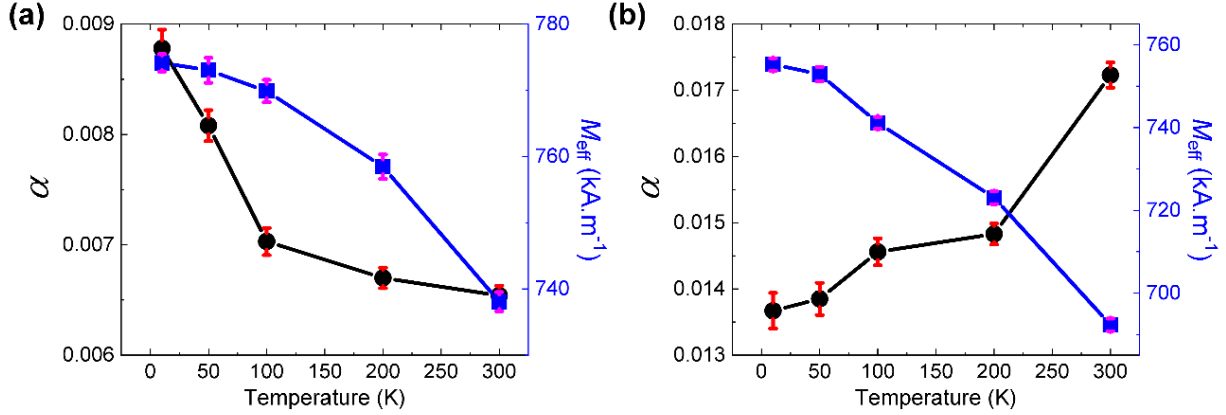

**Figure S5:** Temperature-dependent variation of Gilbert damping parameter($\alpha$) and effective saturation magnetization ($M_{eff}$) in the (a) Py (6.6 nm) and (b) RhSi(49.1 nm)/Py (6.6 nm) thin films.

## V. Permalloy thickness-dependent magnetization dynamics

Within an NM/FM heterostructure, apart from the influence of spin pumping, there exists a distinct probability for the dissipation of spin angular momentum, originating from interfacial depolarization and surface irregularities. These phenomena are commonly referred to as "spin-memory loss" (SML) and "two-magnon scattering" (TMS). In the mechanism of SML, the dissipation of spin angular momentum occurs when the atomic lattice at the interface acts as a reservoir for spin states. This may result from the magnetic proximity effect or interfacial spin-orbit scattering, leading to the transfer of spin polarization to the atomic lattice. TMS, on the other hand, occurs when a uniform FMR mode is disrupted, giving rise to the creation of a degenerate magnon with a different wave vector. The non-conservation of momentum in this process can be explained by considering a pseudo-momentum derived from internal field irregularities or secondary scattering events. Both SML and TMS can significantly contribute to the enhancement of the $\alpha$. In the presence of these effects, the modulation of $\alpha$ can be approximated as follows [S4]:

$$\Delta\alpha = g\mu_B \frac{g_{\uparrow\downarrow}^{eff} + g_{SML}}{4\pi t_{Py} M_s} + \beta_{TMS} t_{Py}^{-2} \qquad (S3)$$

Here, $g_{SML}$ represents the "spin-memory loss conductance" and $\beta_{TMS}$ is the "TMS coefficient". We have employed this equation to analyze the dependence of $\alpha$ on the inverse of Py layer thickness to discern the individual influences of the SML, TMS, and spin pumping (SP). For this study, we utilized samples with and without 43.5 nm RhSi underlayer where Py thickness is varied from 6.6 nm to 30 nm.

Figures S6(a) and S6(b) depict the Py-thickness-dependent variations of $\alpha$ at 300K and 10 K system temperature. In the absence of RhSi, $\alpha$ remains small and independent of Py thickness. However, in the presence of RhSi, $\alpha$ exhibits a linear increase with the inverse of Py thickness. Table S1 presents the values of $\beta_{TMS}$, $g_{SML}$, and $g_{\uparrow\downarrow}^{eff}$ obtained through the fitting of Py-thickness-dependent $\alpha$ using Equation (S3). It shows that, as the temperature decreases, the relative extrinsic contributions diminish significantly. Figure S6(c) shows the Py-thickness-dependent relative contributions of extrinsic TMS and SML effects to $\alpha$ at 300K. It is evident that the extrinsic contribution is well below 10 % of the damping modulation for all the RhSi/Py samples. At lower thickness, this contribution becomes slightly higher but remains much lower than the intrinsic SP contribution. We have extracted the interfacial magnetic anisotropy energy density ($K_s$) which is an indicator of the strength of the interfacial spin-orbit coupling (SOC) by fitting the Py-thickness-dependent $M_{eff}$ (see Figure S6(d)) with the formula [S5]:

$$4\pi M_{eff} = 4\pi M_S - \frac{2K_s}{M_s t_{Py}} \qquad (S4)$$

From the fit, we have extracted the values of $K_s$ to be $77 \times 10^{-5}$ J/m$^2$ and $43 \times 10^{-5}$ J/m$^2$ in the presence and absence of the RhSi underlayer, respectively. This increase in $K_s$ value in the presence of RhSi is an indication of the increase in interfacial SOC strength [S5]. The saturation magnetization ($M_s$) also decreases from 848 kA/m in the absence of RhSi to 813 kA.m$^{-1}$ in the presence of RhSi underlayer. Table S1 presents the temperature-dependent values for $K_s$ and $M_s$. This illustrates that, in accordance with Bloch's law, as the temperature decreases, $M_s$ undergoes an increase, while $K_s$ demonstrates a minor decrease, signifying a subtle enhancement in interfacial SOC strength.

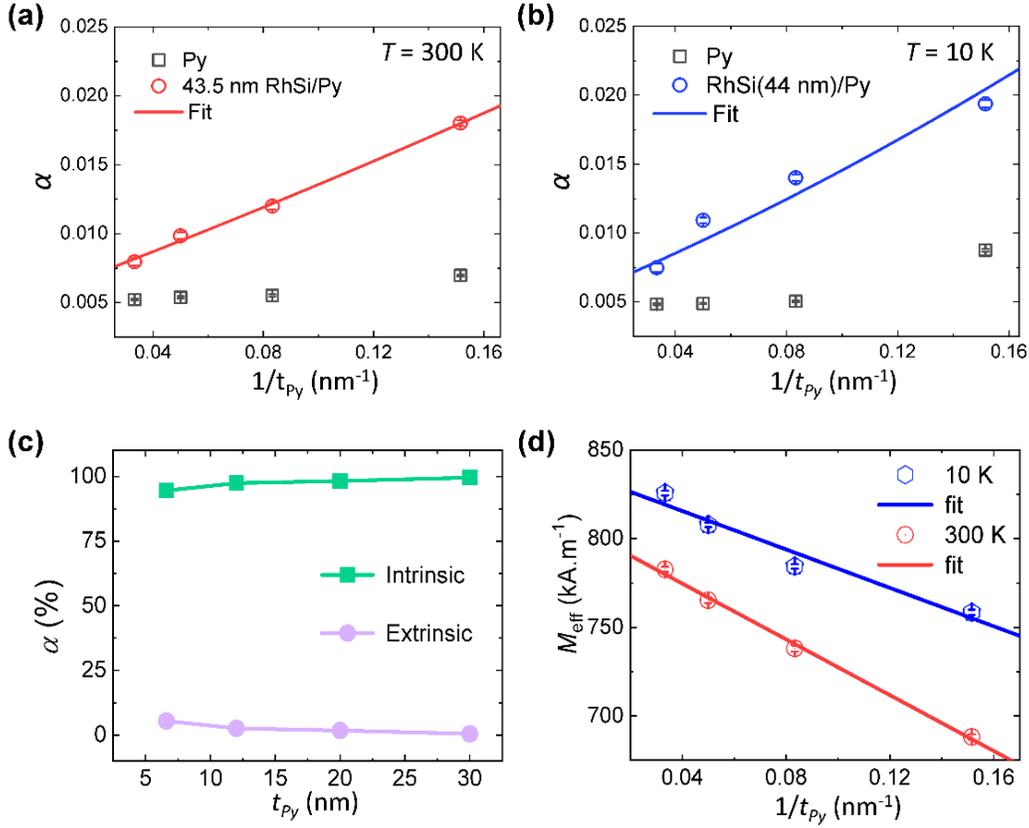

**Figure S6:** Enhancement of effective Gilbert damping parameter with the inverse of permalloy thickness at (a) 300 K and (b) 10 K. The red solid lines represent the fits using Equation (S3). (c) Percentage contribution in the modulation of damping from extrinsic spin memory loss and two-magnon scattering across the thickness range of the permalloy layer. (d) Variation of $M_{eff}$ with $1/t_{Py}$. Symbols are experimental data points and solid red lines are fits using Equation (S4).

**Table S1:** Temperature-dependent spin-mixing conductance ($g_{\uparrow\downarrow}$), spin-memory loss (SML) conductance ($g_{SML}$), two-magnon scattering (TMS) coefficient ($\beta_{TMS}$), saturation magnetization ($M_s$), and interfacial magnetic anisotropy energy density ($K_s$) in the RhSi(43.5 nm)/Py heterostructures.

| Temperature (K) | $\beta_{TMS}$ (nm$^2$) | $g_{SML}$ (nm$^{-2}$) | $g_{\uparrow\downarrow}^{eff}$ (nm$^{-2}$) | $M_s$ (kA.m$^{-1}$) | $K_s$ (10$^{-5}$ J/m$^2$) |
|---|---|---|---|---|---|
| 300 | 0.42 | 1.47 | 38.7 | 813.1 | 43 |
| 10 | 0.18 | 1.01 | 25.3 | 844.2 | 32 |

## Supplementary References